\documentclass{elsart}

\usepackage{epsfig}

\usepackage{amssymb}
\journal{Physics Letters B}

\begin{document}

\begin{frontmatter}



\title{Sub-mm gravity: confronting the modified dynamics with higher-dimensional theories}


\author{S.O. Mendes},
\ead{smendes@iagusp.usp.br}
\author{R. Opher}
\ead{opher@orion.iagusp.usp.br}
\address{Instituto Astron\^omico e Geof\'\i sico, Universidade de S\~ao Paulo, \\
Av. Miguel Stefano 4200, S\~ao Paulo, 04301-904 SP, Brazil}

\begin{abstract}
We propose that future experiments aiming at the detection of deviations from the $1/r^2$ 
gravitational law on submillimetric scales can be used to test the modified Newtonian 
dynamics theory (MOND). Current experiments are able to test the gravitational field of 
masses $m\approx$ 1g at distances $r \approx$ 200 $\mu$m, implying that they are probing 
accelerations well above the MOND limit ($a_0 \approx$ 1.2 $\times$ 10$^{-8}$ cms$^{-2}$). 
We show that MONDian effects begin to be important at the submillimetric level for masses 
$m\leq$ 1 mg. MOND makes predictions that are clearly distinguishable from those expected 
in a scenario with compact extra dimensions. This will enable direct confrontation between 
the two theories if future experiments can improve their mass scales to the milligram 
level.

\end{abstract}

\end{frontmatter}

\section{Introduction}
There exists today an international effort aiming at the detection of deviations from conventional gravity, motivated by the possible existence of new spatial dimensions. The hypothesis of the existence of new spatial dimensions has been proposed as a solution to the hierarchy problem in particle physics. Compact dimensions would make the gravitational interaction appear to be weaker than other interactions, since part of its strength would ``leak'' into other dimensions \cite{AHDD99}. Thus, the gravitational attraction should be stronger on small scales. Some recent gravitational experiments have been searching for exactly this kind of deviation from Newtonian gravity \cite{H01}. We propose in this 
letter that the same experiments can be used to test for a similar prediction of deviations, suggested by the modified Newtonian dynamics (MOND), which is postulated to be valid for accelerations below a certain threshold, $a_0 \approx$ 1.2 $\times$ 10$^{-8}$ cms$^{-2}$. 

We briefly discuss the predicted accelerations for a given mass according to each theory, and then calculate the mass range which makes a direct comparison possible. We also discuss how MOND, in its original form, implicitly violates the strong equivalence principle and suggest that this hypothesis be relaxed. As we shall see, this is a necessary condition for testing MOND effects in the laboratory.

\section{Theories with Extra Dimensions}
The induced Yukawa-type gravitational potential in the context of extra dimensions can be 
written as \cite{H01,KS00}:
\begin{equation}
V(r)=-\frac{GM}{r}(1+\alpha e^{-r/\lambda}),
\label{vsubmm}
\end{equation}
where $\alpha$ and $\lambda$ are the intensity and range of the potential, respectively. 
According to recent experiments, the upper limit for $\lambda$ is of the order of 1 mm. A 
review of theoretical and experimental constraints for both $\lambda$ and $\alpha$ can be 
found in \cite{KS00}. Assuming spherical symmetry, the gravitational acceleration 
corresponding to the above potential is, thus,
\begin{equation}
g_{submm} = \frac{GM}{r}\left[\frac{1}{r}+\alpha 
e^{-r/\lambda}\left(\frac{1}{r}+\frac{1}{\lambda}\right)\right].
\label{gsubmm}
\end{equation}

\section{Modified Newtonian Dynamics}
The dark matter problem on galactic and extragalactic scales has led to the development of 
MOND \cite{M83}. This theory postulates that for accelerations below a certain threshold 
(determined empirically to be $\approx 1.2\times10^8$ cm s$^{-2}$), Newtonian dynamics 
should be no longer valid. It was proposed, instead, that the correct form of the 
gravitational acceleration be given by
\begin{equation}
g = \frac{g_N}{\mu(g/a_0)},
\label{gmond}
\end{equation}
where $g_N$ is the usual Newtonian acceleration and $\mu(x)$ is a function which obeys the 
relation
\begin{equation}
\mu(x) = \left\{ \begin{array}{ll}
                  1,     & \mbox{  $x \gg 1$} \\
                  x,     & \mbox{  $x \ll 1$}
                \end{array}
        \right.
\end{equation}		
A commonly used function having the required asymptotic behavior is \cite{M83}
\begin{equation}
\mu(x) = \frac{x}{\sqrt{1+x^2}}.
\label{mu}
\end{equation}

Unfortunately, MOND cannot be considered to be a complete theory, since a relativistic 
theory whose weak field limit yields MOND does not exist. Some attempts to find a general 
gravitational theory have been made \cite{BM84}, but none of them were completely 
satisfactory. However, MOND fits rotation curve data from spiral galaxies very well with 
only one free parameter, while dark matter fits usually demand three free parameters (see 
\cite{SV98} for a discussion). MOND also predicts the observed Tully-Fisher relation 
\cite{TF77} for spiral galaxies (e.g.~\cite{SV98,S96}), as well as the Faber-Jackson 
\cite{FJ76} relation for ellipticals, as long as M/L does not vary much with mass 
\cite{M84}. The theory has also been applied to other astrophysical phenomena, such as the 
stability and warp of disk galaxies \cite{BM99}, the internal structure of satellite 
galaxies \cite{MO00}, the fundamental plane of elliptical galaxies \cite{S00}, and 
structure formation \cite{S98}.

In addition to the absence of a relativistic theory which incorporates 
MOND, there is yet another difficulty to be dealt with. MOND does not appear to obey the 
strong equivalence 
principle (SEP). Milgrom \cite{M83} suggested, in a seminal paper, that the dynamics of a 
subsystem $s$ of a system $S$ should not be described by MOND when $a_s \ll a_0$ and $a_S 
\gg a_0$ (where $a_s$ and $a_S$ are the typical accelerations in $s$ and $S$, 
respectively). $S$ and $s$ could be, for example, a cluster of galaxies and a single 
galaxy belonging to this cluster, respectively. Thus, when $a_s \ll a_0$ and $a_S \ll 
a_0$, the accelerations are given by MOND, whereas when  $a_s \ll a_0$ and $a_S \gg a_0$ 
they are not given by MOND, but rather by conventional Newtonian theory. This means that 
an observer inside an elevator in free fall, which is embedded in an external homogeneous 
gravitational field, would be able to detect this field. This obviously is a violation of 
the SEP and, if correct, would rule out any attempt of measuring MOND effects in the 
laboratory. Even if one could isolate a system where accelerations are well below $a_0$, 
the strong external fields from the Earth, Moon, and the Sun would erase any MOND 
signature, according to Milgrom \cite{M83}. 

The main motivation for introducing such a violation of the SEP apparently comes from the 
interpretation of data on open stellar clusters in the Solar Neighborhood. According to Jones 
\cite{J70}, the dynamically deduced masses of the clusters Pleiades and Praesepe are about 
1.5 times as large as the mass that can be accounted for by the stars in these clusters. 
The fact that: (1) the internal accelerations within these clusters are a few times smaller 
than $a_0$ (which would lead to a much larger mass discrepancy according to MOND); and (2) 
the accelerations of the clusters produced by the Galaxy is of the order of $a_0$; led 
Milgrom to propose that MOND does not obey the SEP. A sample of 
only two open clusters seems to be too small to be used as a firm indication that MOND 
violates the SEP. A larger sample of both isolated and non-isolated star clusters could 
set the stage for a more complete study of the issue. More recent studies on wide binaries 
\cite{CRC90} and open clusters \cite{LM89} also yield dynamical masses consistent with the 
stellar content. Newer and more refined data might help in this sense (e.g. 
\cite{ESA97}). In this work we neglect the problem of the violation of the SEP in MOND and 
examine the possibility of testing relations (\ref{gmond}) and (\ref{mu}) in the 
laboratory.

\section{MOND $\times$ Extra Dimensions}
\begin{figure}
\vspace{16cm}
\includegraphics{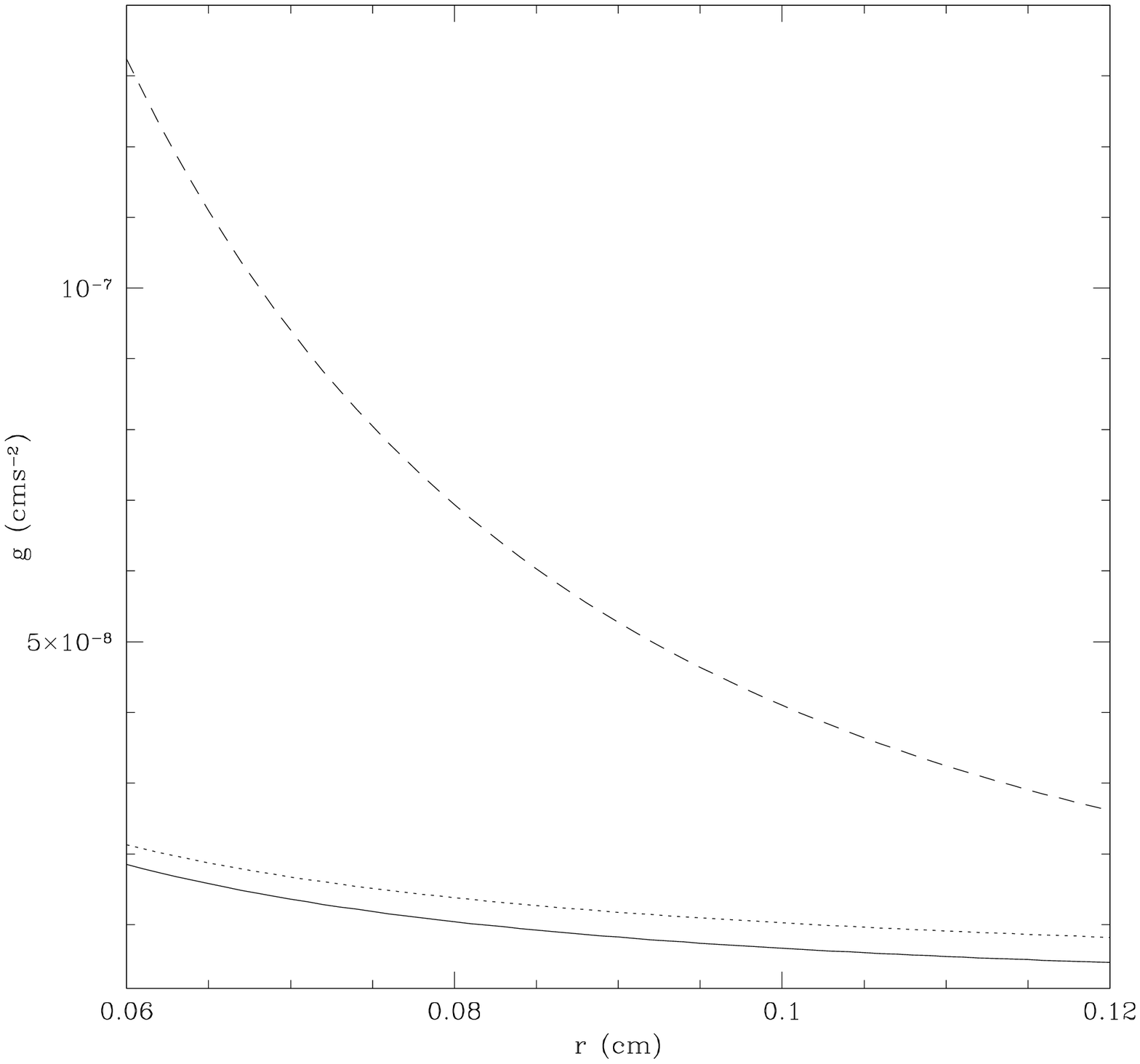}
\caption{Gravitational acceleration as a function of distance produced by a test particle of mass 1 mg according to Newtonian theory (solid), MOND (dotted) and a higher dimensional theory (dashed).}
\label{fig1} 
\end{figure}

Test particles with masses small enough to make MONDian effects noticeable on 
submillimetric scales need to be studied. With the aid of eq.~(\ref{mu}) and taking $g_N$ 
as $GM/r^2$, assuming spherical symmetry, we can solve eq.~(\ref{gmond}) for $g$:
\begin{equation}
g=\frac{1}{r^2}\left[\frac{GM}{2}(GM+\sqrt{G^2M^2+4r^4a_0^2})\right]^{1/2}.
\end{equation}
In order to estimate the limiting mass for which MOND predicts deviations from the $1/r^2$ 
law, we must find the mass M which satisfies
\begin{equation}
g=a_0.
\end{equation}
For example, $M\approx$ 1 miligram when $r=$ 1 mm. Current experiments on sub-mm gravity 
are able to test the gravitational field produced by test-particles with $M\approx$ 1 
gram. Fig.~(\ref{fig1}) illustrates the type of deviations expected from both MOND and 
theories with extra dimensions. We used the ($\lambda-\alpha$) diagram of \cite{KS00} for 
eq.~(\ref{gsubmm}) and assumed that $\lambda\approx$ 1 mm and $\alpha\approx$ 7. If future 
experiments succeed in testing the required MOND mass scale at the sub-mm level, we will 
then be able to test for the presence of MONDian effects in strong external gravitational 
fields. MOND makes predictions that are clearly distinguishable from those of theories 
with new compact dimensions, which makes them competitive in interpreting future 
experimental results. The major problem is, of course, that very small masses are required 
for a direct comparison. We leave the question of whether such an achievement is plausible 
in the near future to the experimentalists. MOND could also be tested in other 
regions of the mass-distance parameter-space: (60 g, 1 cm), (1 kg, 1 m), etc. In this 
case, one would test MOND alone, since this is well beyond the range expected for the 
strengthening of gravity due to extra dimensions.

\section*{Acknowledgements}
The authors are grateful to C.D. Hoyle who kindly clarified some aspects of the experiment 
discussed in \cite{H01}. SOM would like to thank the Brazilian agency CAPES for 
financial support. RO would like to thank the Brazilian agencies FAPESP, CNPq and 
PRONEX/FINEP (no. 41.96.0908.00) for partial support. The authors also thank FAPESP for partial
support through project no. 2000/06770-2.

\clearpage

\end{document}